\documentclass[twocolumn, showpacs,preprintnumbers,nofootinbib,prd,superscriptaddress,10pt]{revtex4-1}
\usepackage{graphicx}
\usepackage{xcolor}
\usepackage{lipsum}
\usepackage{amsmath,amssymb}

\begin{document}
\title{Quantum Dynamics of the Isotropic Universe in the Metric $f(R)$ Gravity}
\author{Mariaveronica De Angelis}
\email{deangelis.1644941@studenti.uniroma1.it}
\affiliation{Physics Department (VEF), Sapienza University of Rome, P.le A. Moro 5 (00185) Roma, Italy}

\author{Laria Figurato}
\email{figurato.1230833@studenti.uniroma1.it}
\affiliation{Physics Department (VEF), Sapienza University of Rome, P.le A. Moro 5 (00185) Roma, Italy}

\author{Giovanni Montani}
\email{giovanni.montani@enea.it}
\affiliation{Physics Department (VEF), Sapienza University of Rome, P.le A. Moro 5 (00185) Roma, Italy}
\affiliation{ENEA, Fusion and Nuclear Safety Department, C.R. Frascati, Via E. Fermi 45 (00044) Frascati (RM), Italy}
	
\begin{abstract}
We analyse the canonical quantum dynamics of the isotropic universe, as emerging from the Hamiltonian formulation of a metric $f(R)$ gravity, viewed in the \emph{Jordan frame}.
The canonical method of quantization is performed by solving the Hamiltonian constraint before quantizing and adopting like a relational time the non-minimally
coupled scalar field emerging in the \emph{Jordan frame}.
The resulting Schr\"odinger evolution is then investigated both in the vacuum and in the presence of a massless scalar field, though as the kinetic component of an inflaton.
We show that, in vacuum, the morphology of localized wave packets is that of a non-spreading profile up to the cosmological singularity. When the external scalar field is included into the dynamics, we see that the wave packets acquire the surprising feature of increasing localization of the universe volume, as it expands.
This result suggests that, in the metric $f(R)$ formulation of gravity, a spontaneous mechanism arises for the Universe classicalization. Actually,
when the phase space of the scalar field is fully explored, such an increasing localization in the Universe volume is valid up to a given value of the time, \emph{i.e.} of the non-minimally coupled mode after which the wave packets spread again. We conclude our analysis by inferring that before this critical transition age is reached, the inflationary phase could take place, here modelled via a cosmological constant. This point of view provides an interesting scenario for the transition from a Planckian Universe to a classical de-Sitter phase, which in the $f(R)$ gravity appears more natural than in the Einsteinian picture.
\end{abstract}

\maketitle
\section{Introduction}

Among the extended (non-Einsteinian) theories of gravity \cite{od}, the so-called $f(R)$-model \cite{SF,C, Nojiri:2017ncd} is one of the most studied in view of possible phenomenology for new physics, like dark energy and dark matter \cite{N}. The reasons for such a wide implementation of this approach to generalize the Einstein-Hilbert action
\cite{Primordial} are to be identified in the simplicity of the reformulation for the generalized Einstein equations, now becoming of the forth order in differentiation.
Actually, suitable expansions of the $f(R)$ in specific limits (e.g. for large or small values of the scalar of curvature $R$), allow a straightforward comparison with the phenomenology of General Relativity, favouring tests of validity for the revised scenario \cite{Zakarov, io e Lecian}.

However, the value of the $f(R)$ formulation for searching new physics is evident when the scalar-tensor setup of this approach is
outlined \cite{SF, Primordial}. Actually, the scalar degree of freedom contained in the form of the function $f$ can be translated into the dynamics of a  scalar field, non-minimally coupled to gravity in the so-called \emph{Jordan frame}, while it can be restated in a minimally coupled form in a formulation dubbed the \emph{Einstein frame} (involving a conformal re-scaling of the metric tensor).

Here, we provide an application of the $f(R)$ model in the \emph{Jordan frame} to quantum cosmology (for applications to classical cosmology see
\cite{bcm, Lattanzi, Moriconi, N, io e Lecian} and for the most reliable models accounting for dark energy see also
\cite{HS, staro, tsu}).

In particular, we consider the evolution of an isotropic universe in the presence of an external minimally coupled scalar field in addition to the intrinsic non-minimally coupled one of the \emph{Jordan frame}.
Actually, the non-minimally coupled mode has always a potential term, summarizing the morphology of the original $f$ (for a gauge-invariant
description of the theory degrees of freedom see
\cite{Moretti-Bombacigno-io}) which is reliably negligible near enough to the initial singularity since its relevance is controlled by the universe volume.

The specific point of view we address here is that one in which the $f(R)$ induced scalar mode plays the role of a relational clock for the canonical quantum dynamics.
This scenario was firstly investigated in \cite{bcm} on the context of the Wheeler-Dewitt equation while here we consider an approach based on an ADM-reduction \cite{ADM62} which leads to a Schr\"odinger-like formulation of the quantum evolution.
In other words, we chose here the non-minimally coupled scalar field as a time variable before quantizing the system \cite{Isham92}.

We firstly analysed the vacuum theory and we see that the dynamics of a localized wave packet remains non-spreading for all the system evolution, since the average values follow classical trajectories which reach the singularity.

Then, we introduce an external matter, in the form of a massless scalar field, to be thought as the kinetic component of an inflation field \cite{Primordial, articolo con Muccino} and, as first step, the simplified case when this scalar field has a small conjugate momentum (\emph{i.e.} it is limited to explore a small phase space region).

In this case, we observe a very peculiar feature, concerning a progressive localization of the wave packet from the singularity toward the fully expanded universe. In other words, we observe a spontaneous mechanism of "classicalization'' of the universe as the relational time flows.
To confirm the validity of this result also in an extended context, we then analyse the general case, which include the possibility of a second regime in the quantum solution, due to the negative character of the argument of a square root present in the Hamiltonian (prevented when the phase space is restricted).
The presence of this new regime does not completely suppress the classicalization phenomenon, which is still present up to a given instant of
time after which, the change of sign of the mentioned square root induces a new spreading tendency of the wave packet.

It is easy to realize how the present analysis is of relevant interest in primordial cosmology, since it offers a natural scenario via the $f(R)$ gravity, as viewed in the \emph{Jordan frame}, to account for the
classical nature of the isotropic Universe, as emerging from a Planckian era \cite{Primordial, Canonical, grav, kt}.

It is worth nothing that classicalization phenomenon is of interest for the real history of a primordial universe and in this respect, the following two perspectives can be considered: 
\begin{itemize}
	
\item The phase space of the inflaton field remains small from the Planck era up to the inflation age
\item The evolution of the primordial universe is affected by new physics before the critical instant, when it starts to de-localize (it is reached).
\end{itemize}

The first hypothesis can be supported only if we consider suitable initial conditions and it calls attention for further investigation when the radiation contribution of the thermal bath is included.
The second scenario appears the most natural since the new physics is naturally identified in the phase transition at the beginning of the inflation (say a huge
cosmological constant term arises) which can prevent the subsequent spreading of the universe wave function.
Here, we infer that when the Universe is
characterized by a very peaked wave function, de facto it is a classical universe, the evolution is naturally driven toward a de-Sitter phase via the emergence of the cosmological constant (no spreading has time to occur). 

The present manuscript is structured as follows.\\
In Sec.(\ref{II}), we present the modified theory of gravity $f(R)$ and construct the Lagrangian formalism in scalar-tensor representation, making particular attention to the \emph{Jordan frame} in which the $f(R)$ gravity is expressed through a non-minimally coupled scalar field to gravity. In Sec.(\ref{III}) we extend the method of Sec.(\ref{II}) in order to derive the Hamiltonian formalism of gravity in the case of $f(R)$ theories in scalar-tensor representation for the isotropic universe. In Sec.(\ref{IV}) we perform a critical study on the canonical quantization of the isotropic universe in the case of the $f(R)$ theories of gravity in the \emph{Jordan frame}. In particular, using the scalar field $\xi$ as the pre-quantization time and developing an analysis on the quantum dynamics of the FLRW model in the vacuum. In Sec.(\ref{V}) we repeat the analysis of Sec.(\ref{IV}) adding the matter scalar field $\phi$. We perform this study considering two different cases: the first one within a perturbation approach in order to facilitate calculations and so to restrict the dynamics of the field in a small region of the phase space, and the second one regarding its exact quantum evolution. Moreover, we highlight also the classical behaviour and we discuss the role of inflation in constructing a reliable classicalization of the universe. Finally, in Sec.(\ref{VI}) conclusions are drawn.

\section{f(R) theories in the Jordan frame}\label{II}
Einstein's theory of General Relativity satisfactory describes gravity phenomenology. Even if the determination of the gravitational field kinematics is a very consistent formulation due to its geometrical and tensorial structure, a wide class of different proposals are admitted by its dynamics. Actually, the Einstein-Hilbert action contains the Ricci scalar $R$ that is only the most simple proposal. In fact, the Ricci scalar depends on the second derivatives of the metric tensor $g_{\mu \nu}$ and, under variation, it provides a second order differential equations. These metric $f(R)$ theories come from a straightforward generalization of the Lagrangian in the Einstein-Hilbert action as follows
\begin{equation}
S_{EH}=-\frac{1}{2 \kappa}\int d^4x \sqrt{-g}f(R),
\label{action}
\end{equation}
where $\kappa \equiv \frac{8\pi G}{c^4}$, $G$ is the gravitational constant, $g$ is the determinant of the metric and the function $f$ corresponds to $\infty^1$ degrees of freedom. 
In order to derive the Einstein's field equations, in literature there are two possible variational principles: affine and metric approach and in this study, we will refer to the last one. 

We can construct the total action for $f(R)$ gravity by adding a matter term $S_M$,
\begin{equation}
S=-\frac{1}{2 \kappa}\int d^4x \sqrt{-g}f(R)+S_M(g_{\mu \nu}, \psi),
\label{actionm}
\end{equation}
in which $\psi$ represents a generic matter field.

The variation of the action (\ref{action}) with respect to the metric $g_{\mu \nu}$ implies the following set of covariant field equations, characterized by fourth-order differentiation
\begin{equation}
f(R)'R_{\mu \nu}-\frac{1}{2}f(R)g_{\mu \nu}-[\nabla_{\mu} \nabla_{\mu}-g_{\mu \nu} \Box ]f'(R)=\kappa T_{\mu \nu},
\end{equation}
where $f'(R) \equiv df/dR$ and $\Box \equiv g^{ij}\nabla_i\nabla_j$ is the d'Alambert operator in curved manifolds. In order to rewrite both the action and equations in a more convenient way, one can introduce two auxiliary fields (\emph{i.e.} Lagrange multipliers) $A$ and $B$, so the action takes the form
\begin{equation}
S=-\frac{1}{2 \kappa}\int d^4x \sqrt{-g}[B(A-R)+f(A)]+ S_M(g_{\mu \nu}, \psi).
\label{actionf}
\end{equation}
The variation with respect to $B$ gives $R=A$, while the variation with respect to $A$ provides $B=-df/dA \equiv -f'(A)$. We now redefine the field $A$ by introducing the scalar field $\xi=f'(A)$ and its potential 

\begin{equation}
V(\xi)=A(\xi)\xi-f(A(\xi)).
\label{potential}
\end{equation}
In this way the action (\ref{actionf}) is now rewritten as
\begin{equation}
S_{J}=-\frac{1}{2 \kappa}\int d^4x \sqrt{-g}[\xi R-V(\xi)]+S_M(g_{\mu \nu}, \psi),
\label{actionj}
\end{equation}
and it is known as the $f(R)$ gravitational action in the \emph{Jordan frame}. Thus, we get a scalar-tensor representation based on translating the scalar degree of freedom related to the function $f(R)$ into a dynamical scalar field non-minimally coupled with the curvature $R$. 

Hence, the field equations turn out to be
\begin{equation}
\begin{cases}
G_{\mu \nu}=\frac{\kappa}{\xi}T_{\mu \nu}-\frac{1}{2 \xi}g_{\mu \nu}V(\xi)+\frac{1}{\xi}(\nabla_{\mu}\nabla_{\nu}\xi-g_{\mu \nu}\Box \xi)=\kappa T\\R=V'(\xi)
\label{secondorder}
\end{cases}
\end{equation}
corresponding now to a second-order formulation. Taking the trace of the first equation (\ref{secondorder}) and using the second one we get
\begin{equation}
3\Box \xi +2V(\xi)-\xi\frac{dV}{d\xi}=\kappa T
\end{equation}
that determines the dynamics of the scalar field for a given source of matter.

\section{Hamiltonian formulation of the isotropic Universe dynamics}\label{III}

We start considering a closed (positive curved) FLRW model which corresponding ADM line element \cite{kt} is provided by the expression
\begin{equation}
ds^2=g_{\mu \nu}dx^{\mu}dx^{\nu},
\end{equation}
in which the metric tensor $g_{\mu \nu}$ is written as
\begin{equation}
  g_{\alpha\beta} =
\begin{pmatrix}
N^2 & 0 & 0 & 0 \\
0 & -a^2 & 0 & 0 \\
0 & 0 & -a^2 sin^2\chi & 0 \\
0 & 0 & 0 & -a^2 sin^2\chi sin^2\theta
\end{pmatrix}. 
\end{equation}
In this way, the covariant volume element necessary for the construction of the ADM gravity action is the following
\begin{equation}
\sqrt{-g}=Na^3sin^2\chi sin\theta.
\label{-g}
\end{equation}
Using the $f(R)$ gravity action in the scalar-tensor formulation of the \emph{Jordan frame} (\ref{actionj}) and outlining the contribute of $\xi R$ as
\begin{equation}
    \xi R = -6 \xi \Bigg\{ \Big[ \frac{1}{aN} \frac{\partial}{\partial t} \Big(\frac{1}{N} \frac{\partial a}{\partial t} \Big) \Big] + \Big[ \frac{1}{a^2N^2} \Big( \frac{\partial a}{\partial t} \Big)^2 \Big] + \frac{1}{a^2} \Bigg\},
\end{equation}
the action takes the form

\begin{equation}
S_{JF} ^{ADM}= \frac{3\pi c^3}{8G} \int dt \Big(-\frac{\xi a \dot{a}^2}{N} - \frac{a^2 \dot{a} \dot{\xi}}{N} +Na \xi +\frac{a^3 N V(\xi)}{6} \Big).
\end{equation}
Hence, the ADM Lagrangian density reads

\begin{equation}
     \mathcal{L}_{JF} ^{ADM}=- \frac{3\pi c^3}{4G} \Big( \frac{\xi a \dot{a}^2}{N} + \frac{a^2 \dot{a} \dot{\xi}}{N} -Na \xi -\frac{a^3 N V(\xi)}{6} \Big).
\label{lagrangian}
\end{equation}

In order to pass to the Hamiltonian formulation of gravity, let us perform a Legendre transformation by defining the momentum $p_a$ conjugated to the scale factor $a$ and the momentum $p_{\xi}$ conjugated to the variable $\xi$ as
\begin{equation}
 p_a 
 = - \frac{3 \pi c^3}{4 G N } (2 \xi a \dot{a} + a^2 \dot{\xi} )   
 \label{pa}
\end{equation}
\begin{equation}
  p_\xi 
  = - \frac{3 \pi c^3}{4 G N}  a^2 \dot{a} .  
  \label{pxi}
\end{equation}
In this way we get
\begin{equation}
 \dot{a}=-\frac{4GN}{3\pi c^3}\frac{p_\xi}{a^2},
\label{ap}
\end{equation}

\begin{equation}
 \dot{\xi} = \frac{8GN}{3 \pi c^3} \frac{\xi}{a^3} p_\xi - \frac{4GN}{3 \pi c^3} \frac{p_a}{a^2}.
\label{xip}
\end{equation}
So that, the Hamiltonian function 
\begin{equation}
\mathcal{H}_g\equiv p_a \dot{a}+p_ {\xi}\dot{\xi}-\mathcal{L}_g
\label{Hd}
\end{equation}
is obtained using (\ref{lagrangian}), (\ref{ap}) and (\ref{xip}) as follows

\begin{equation}
    \mathcal{H}_g = N \biggl(- \frac{4G}{3 \pi c^3} \frac{p_a p_\xi}{a^2} + \frac{4G}{3 \pi c^3} \frac{\xi}{a^3} p_\xi^2 - \frac{3 \pi c^3}{4G} a \xi - \frac{\pi c^3}{8G}  a^3 V(\xi)\biggl).
\label{Hg}
\end{equation}

In order to introduce the matter component with a free scalar field which in a cosmological setting depends only on time $\phi(t)$, we consider a matter action and (\ref{-g}) as follows
\begin{equation}
    S_\phi = \frac{2\pi ^2}{c} \int dt  \biggl(\frac{a^3}{2N} \dot{\phi}^2 -a^3NV(\phi)\biggl).
    \label{matteraction}
\end{equation} 

Then, performing the Legendre transformation by defining the momentum $p_{\phi}$ conjugate to the variable $\phi$ as
$$
p_\phi
= \frac{2 \pi^2}{c} \frac{a^3}{N} \dot{\phi}
$$
\begin{equation}
\rightarrow \dot{\phi} = \frac{c N}{2 \pi^2} \frac{p_\phi}{a^3}
\label{pphi}
\end{equation}
and following the same steps as before, we achieve

\begin{equation}
\mathcal{H}_\phi = N\biggl(\frac{c}{4 \pi^2} \frac{1}{a^3} p_\phi^2 + \frac{2 \pi^2}{c} a^3 V(\phi)\biggl).
\label{Hphi}
\end{equation}
Hence, the total gravity-matter \emph{super-Hamiltonian} reads as 
\begin{widetext}
	\begin{equation}
     \mathcal{H} = \frac{4G}{3 \pi c^3} \frac{\xi}{a^3} p_\xi^2 + \frac{c}{4 \pi^2} \frac{1}{a^3} p_\phi^2 - \frac{4G}{3 \pi c^3} \frac{p_a p_\xi}{a^2}-\frac{3 \pi c^3}{4G} a \xi-  a^3 \Big[ \frac{\pi c^3}{8G} V(\xi) - \frac{2 \pi^2}{c} V(\phi) \Big].
\label{Htot}
\end{equation}
\end{widetext}

It is important to stress that variating the total action with respect to $N$ we easily get that the gravity-matter \emph{super-Hamiltonian} above must vanish.

\section{Schr\"odinger dynamics in vacuum case}\label{IV}

In this section, we extend the \emph{minisuperspace} formalism to the $f(R)$ scalar-tensor representation in which the scalar field $\xi$ mimics the role of the internal time variable \cite{bcm}, chosen before quantizing the system.
We assume that, near the Planckian era (\emph{i.e.} $a\rightarrow 0$), we can neglect the potential terms as they are dominated by a factor $a^3$ as well as the linear term $a$ due to the positive spatial curvature. In order to study the vacuum case, we set $p_{\phi}\equiv 0$. Hence, the Hamiltonian is written as
\begin{equation}
     \mathcal{H}
     =  \frac{4GN}{3 \pi c^3} \Bigg( \frac{\xi}{a^3} p_\xi^2 
     - \frac{p_a p_\xi}{a^2} \Bigg).
    \label{Ha}
\end{equation}
It is now convenient to adopt the change of variable $\alpha=\ln a^2$. Equation (\ref{Ha}) takes the form 
\begin{equation}
\mathcal{H} = \frac{4GN}{3 \pi c^3} e^{-\frac{3}{2}\alpha} \Bigg( \frac{1}{2} \xi p_\xi^2 
- p_\alpha p_\xi \Bigg).
\label{Halpha}
\end{equation}


According to the ADM procedure \cite{1962PhRv..126.1864B, grav}, we solve the \emph{super-Hamiltonian} constraint with respect to the conjugate momentum to the time-like variable $\xi$
$$
\mathcal{H} = \frac{4GN}{3 \pi c^3} e^{-\frac{3}{2}\alpha} \Bigg( \frac{1}{2} \xi p_\xi^2 - p_\alpha p_\xi \Bigg) = 0
$$
\begin{equation}
\rightarrow \  p_\xi \Bigg( \frac{1}{2} \xi p_\xi - p_\alpha \Bigg) = 0,
\end{equation}
hence two possible solutions can be obtained
\begin{equation}
    \begin{cases}
        p_{\xi, 1} = 0\\
        p_{\xi, 2} = \frac{2p_\alpha}{\xi}.
    \end{cases}
\end{equation}
Choosing the non trivial solution we fix the Hamiltonian that rules the dynamics of the theory with respect to $\xi$
\begin{equation}
h_\xi = -\frac{2p_\alpha}{\xi}.
\end{equation}
The next step consists in setting the \emph{time gauge} $\dot{\xi}=1$ that fixes the \emph{lapse function} $N$ to the form
\begin{equation}
    N_{ADM} = \frac{3 \pi c^3}{4G} \frac{e^{-\frac{3}{2}\alpha}}{\Big( \xi p_\xi - p_\alpha \Big)}
\end{equation}
hence we can the reduced ADM action as follows
\begin{equation}
    S_{red} = \int d\xi \Big( p_\alpha \frac{\partial \alpha}{\partial \xi} - h_\xi \Big).
    \label{acred}
\end{equation}

According to (\ref{acred}) we derive the Hamilton equations
\begin{equation}
    \begin{cases}

         \frac{\partial \alpha}{\partial \xi} = \frac{\partial h_\xi}{\partial p_\alpha} = -\frac{2}{\xi}\\
         \frac{\partial p_\alpha}{\partial \xi} = -\frac{\partial h_\xi}{\partial \alpha} = 0\\

    \end{cases}
\end{equation}
which determine the classical trajectory with respect to the time $\xi$
\begin{equation}
    \alpha(\xi) = -2ln(\xi) + \alpha_0,
    \label{tc}
\end{equation}
where $\alpha_0$ is set equal to zero in order to have $\alpha(0)=0$.

Finally, promoting $h_\xi$ to a quantum operator we obtain the equation \emph{à la} Schr\"odinger
\begin{equation}
     \frac{\partial \psi}{\partial \xi} = \frac{2 \partial \psi}{\xi \partial \alpha}
    \label{sc}.
\end{equation}
Substituting the plane wave representation $\psi(\alpha, \xi)=e^{i k_{\alpha}\alpha}\gamma(\xi)$ we obtain a differential equation for $\gamma(\xi)$, \emph{i.e.}:
\begin{equation}
    \frac{\partial \gamma}{\partial \xi} = \frac{2ik_\alpha}{\xi} \gamma
\end{equation}
having the solution $\gamma(\xi)=d_1\ \xi^{2i k_{\alpha}}$, where $d_1$ is an integration constant. Hence, we get 
\begin{equation}
    \psi(\alpha, \xi)=d_1 e^{i k_{\alpha} \alpha}\xi^{2 i k_{\alpha}}.
\end{equation}

The next step is to construct a localized wave packet by using a Gaussian distribution on $k_{\alpha}$ (\ref{sc}), \emph{i.e.}:
\begin{equation}
    \Psi(\alpha, \xi)=\int^{+\infty}_{-\infty} dk_{\alpha} A(k_{\alpha})\psi(\alpha, \xi),
\end{equation}
where we assumed
\begin{equation}
    A(k_{\alpha})=\frac{1}{\sqrt{2\pi}\sigma_{\alpha}}e^{-\frac{(k_{\alpha}-\bar{k_{\alpha}})^2}{2\sigma_{\alpha}^2}}.
\end{equation}

\begin{figure}[h!] 
\centering
\includegraphics[width=8.5cm]{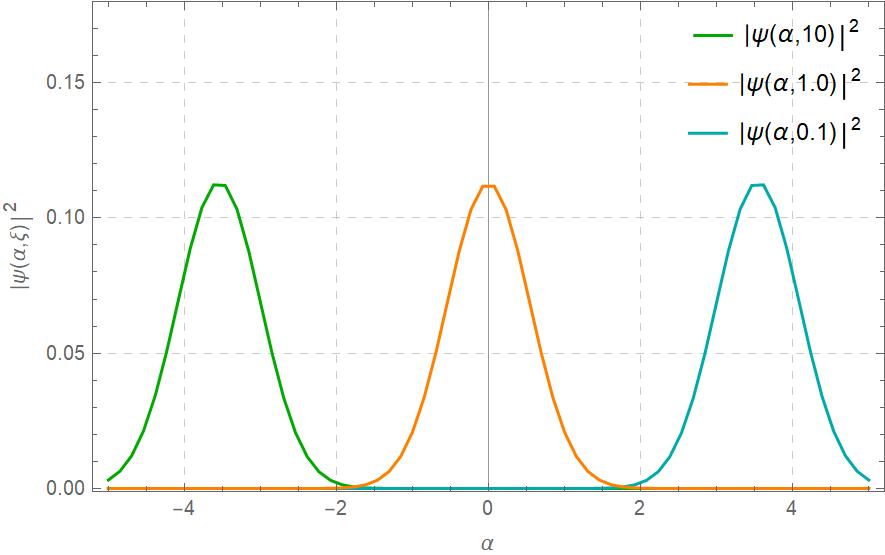} 
\caption{Evolution of the square modulus of the wave function $\Psi(\alpha, \xi)$ for different values of time. We set $\bar{k}_{\alpha}=1$, $\sigma_{\alpha}=1$.}
\label{psiquadroalpha}
\end{figure}

Hence, according to the natural scalar product for a Schr\"odinger-like equation, the normalizable probability density is
\begin{widetext}
\begin{equation}
    |\Psi|^2=\Psi \Psi^*=\int^{+\infty}_{-\infty}dk_{\alpha} A(k_{\alpha})e^{ik_{\alpha}\alpha}\xi^{2ik_{\alpha}}\int^{+\infty}_{-\infty}dk_{\alpha}A(k_{\alpha})e^{-ik_{\alpha}\alpha}\xi^{-2ik_{\alpha}},
    \label{psiquadro}
\end{equation}
\end{widetext}
which provides the probability of finding the universe at a certain instant $\xi$ for unit of spatial coordinate $\alpha$. Therefore, it is possible to evaluate the consistency of the quantum cosmological model under consideration by studying the behavior of quantity (\ref{psiquadro}) as time varies Fig.(\ref{psiquadroalpha}), \emph{i.e.} $\xi > 0$. All the calculations of this work are performed with the help of \emph{Wolfram Mathematica}.

It is important to stress that the probability density remains perfectly localized over time. 

\begin{figure}[h!] 
\centering
\includegraphics[width=8.5cm]{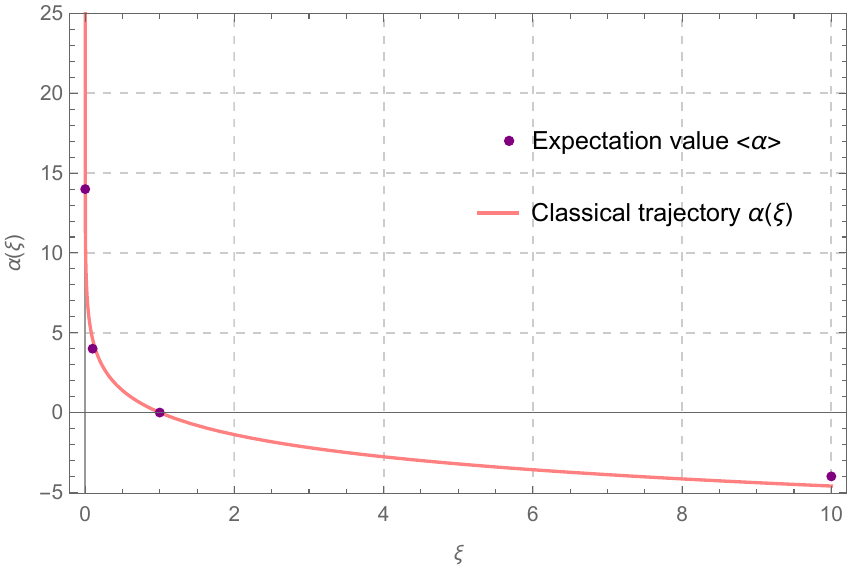} 
\caption{Comparison between the classical trajectory (solid line) and the $\alpha(\xi)$ values where the
probability density is maximally localized (points).}
\label{traiettoriaclassicaalpha}
\end{figure}

To demonstrate the consistency of the classical limit with the quantum behaviour, we compare the classical trajectory (\ref{tc}) provided by the Hamilton equations with the values of the $\alpha$ coordinates corresponding to the peaks of the probability density (\ref{psiquadro}) as $\xi$ varies. In fact, for probability densities well localized, as in the case of a Gaussian, the value of $\alpha$ for which there is a maximum closely resembles the expectation value behaviour. From Fig.(\ref{traiettoriaclassicaalpha}), it is clear that the quantum dynamics formalized by the wave packet $\Psi(\alpha,\xi)$ correctly follows the classical dynamics of the FLRW universe in the \emph{Jordan frame}. 

We highlight the remarkable result of having shown that a cosmological model of this type reaches the singularity ($\xi \rightarrow + \infty$, $\alpha \rightarrow - \infty$) in a quasi-classical way, allowing to consider the quantum effects in the Planckian regime as lower-order effects on a semi-classical Universe.

\section{Quantum dynamics in the presence of matter}\label{V}
In this section, we analyse the Hamiltonian (\ref{Halpha}) considering the free external scalar field $\phi$. Hence, it becomes 

\begin{equation}
\mathcal{H} = \frac{4GN}{3 \pi c^3} e^{-\frac{3}{2}\alpha} \Bigg( \frac{1}{2} \xi p_\xi^2 + \frac{3 c^4}{16 \pi G} p_\phi^2 
- p_\alpha p_\xi \Bigg).
\end{equation}

Thus, solving the \emph{super-Hamiltonian} constraint ($\mathcal{H}=0$)with respect to the conjugate momentum $p_\xi$, it reads as
\begin{equation}
    p_\xi = \frac{p_\alpha \pm \sqrt{p_\alpha^2 - \frac{3c^4}{8\pi G}\ \xi\ p_\phi^2}}{\xi},
\end{equation}
we choose the positive solution in order to achieve that one we found in the case without the scalar field and define the Hamiltonian that rules the dynamics with respect to $\xi$ as
\begin{equation}
    h_\xi = - \Bigg( \frac{p_\alpha + \sqrt{p_\alpha^2 - \frac{3c^4}{8\pi G}\ \xi\ p_\phi^2}}{\xi} \Bigg).
    \label{hximatter}
\end{equation}
Hence, the Hamilton equations are
\begin{equation}
    \begin{cases}
    \frac{\partial \alpha}{\partial \xi}=\frac{\partial h_{\xi}}{\partial p_{\alpha}}=\frac{1}{\xi}\biggl(-1-\frac{2p_{\alpha}}{\sqrt{p_{\alpha}^2-\frac{3c^4}{8\pi G}\xi p_{\phi}^2}}\biggl)\\
    \frac{\partial p_{\alpha}}{\partial \xi}=-\frac{\partial h_{\xi}}{\partial \alpha}=0
    \end{cases}
    \label{hamiltonesatte}
\end{equation}
which determine the classical trajectory

\begin{equation}
    \alpha(\xi)=\alpha_0-3 \log [p_{\alpha}-\sqrt{p_{\alpha}^2-3\xi p_{\phi}^2}]+\log[p_{\alpha}+\sqrt{p_{\alpha}^2-3\xi p_{\phi}^2}]
    \label{alphacampo}
\end{equation}
with $\alpha_0$ an integration constant and $p_{\alpha}$ still a constant of motion.

Promoting $h_\xi$ and $p_{\xi}$ to quantum operators we derive the equation \emph{à la} Schr\"odinger

\begin{equation}
     i \frac{\partial \psi}{\partial \xi} = - \frac{1}{\xi} \Bigg ( - i \partial_\alpha + \sqrt{-  \partial_\alpha^2 + \frac{3c^4}{8\pi G} \xi  \partial_\phi^2} \Bigg ) \psi;
    \label{scroesatta}
\end{equation}
as before, we use the \emph{ansatz} $\psi(\alpha, \phi, \xi)=e^{i( k_{\alpha}\alpha+ k_{\phi}\phi)}\beta({\xi})$ and by applying to the plane wave the square root derived from the non-local Hamiltonian \cite{doi:10.1063/1.530015}, we obtain
\begin{equation}
     i \xi \partial_\xi \beta(\xi) = \Bigg ( -k_\alpha - \sqrt{k_\alpha^2 - \frac{3c^4}{8\pi G} \xi k_\phi^2} \Bigg ) \beta(\xi).
\end{equation}
Now, we consider the change of variable $\xi=e^{\tau}$ in order to achieve an equation for $\beta(\tau)$ as
\begin{equation}
     i \partial_\tau \beta(\tau) = \Bigg ( -k_\alpha - \sqrt{k_\alpha^2 - \frac{3c^4}{8\pi G} k_\phi^2 e^{\tau}} \Bigg ) \beta(\tau).
     \label{betatau}
\end{equation}

In the following subsections, we consider two different cases. The perturbation approach which consists of having a small conjugate momentum of the scalar field (so that the argument of the square root remains always positive) and eventually the general formulation where no approximation is made.

\subsection{Perturbation approach}\label{subsecA}
To facilitate the calculation, we start by restricting the dynamics of the scalar field $\phi$ to a small region of the phase space. In other words, we expand the square root of (\ref{betatau}) with Taylor polinomials, obtaining 
\begin{equation}
    \partial_\tau \beta(\tau) \simeq \biggl(2ik_\alpha - \frac{i}{2} \frac{3c^4}{8\pi G} \frac{k_\phi^2}{k_\alpha} e^{\tau}\biggl)\beta(\tau).
\end{equation}
Hence, an equation for $\beta(\tau)$ as well as for $\beta(\xi)$ is found
\begin{equation}
    \beta(\xi)=\xi^{2i k_\alpha} e^{-\frac{i}{2}\frac{3c^4}{16\pi G}\frac{k_\phi^2}{k_\alpha} \xi},
\end{equation}
and the complete wave function is written as
\begin{equation}
    \psi(\alpha, \phi, \xi)=e^{i( k_{\alpha}\alpha+ k_{\phi}\phi)}\xi^{2i k_\alpha} e^{-\frac{i}{2}\frac{3c^4}{16\pi G}\frac{k_\phi^2}{k_\alpha} \xi}.
\end{equation}

As described in the preceding paragraph, we construct the normalizable probability density (\ref{psiquadro}), associated to a Gaussian wave packet in $k_{\alpha}$ and $k_{\phi}$ (running in a small domain around zero), which gives the probability of finding the universe at a certain instant $\xi$ per unit of spatial coordinates $\alpha$ and $\phi$. 
\begin{figure}[h!] 
\centering
\includegraphics[width=8.5cm]{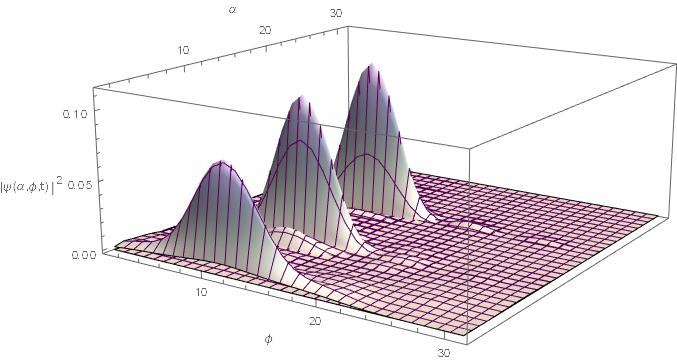} 
\caption{Evolution of the square modulus of the wave function $\Psi(\alpha, \phi, \xi)$ with the perturbation approach. We take respectively $\xi= 10, 1, 0.1$; $\bar{k}_{\alpha}=\bar{k}_{\phi}=0$ and $\sigma_{\alpha}=\sigma_{\phi}=5$.}
\label{tretempi}
\end{figure}
In Fig.(\ref{tretempi}) we observe a very peculiar feature concerning a  progressive  localization  of the  wave  packet from the singularity toward the expanding universe. Thus, we see that a spontaneous mechanism of classicalization emerges.

\subsection{Exact regime}
The interesting result obtained leads us to study the general case in order to see this feature is maintained even in this context for arbitrary values of the the scalar field conjugate momentum. 
Hence, substituting the solution of (\ref{betatau}) in the plane wave, we get the total final expression
\begin{widetext}
\begin{equation}
 \psi(\alpha, \phi, \xi)= e^{i(k_{\alpha}\alpha+ k_{\phi}\phi)}\xi^{i k_{\alpha}} e^{2 i\sqrt{k_{\alpha}^2 - \frac{3c^4}{8\pi G} \xi k_{\phi}^2}} \ e^{-2i k_{\alpha}   arctanh{ \Big(  \frac{\sqrt{k_{\alpha}^2 - \frac{3c^4}{8\pi G}\xi k_{\phi}^2}}{k_{\alpha}}} \Big) }.  
\end{equation}
\end{widetext}

\begin{figure}[h!] 
\centering
\includegraphics[width=8.5cm]{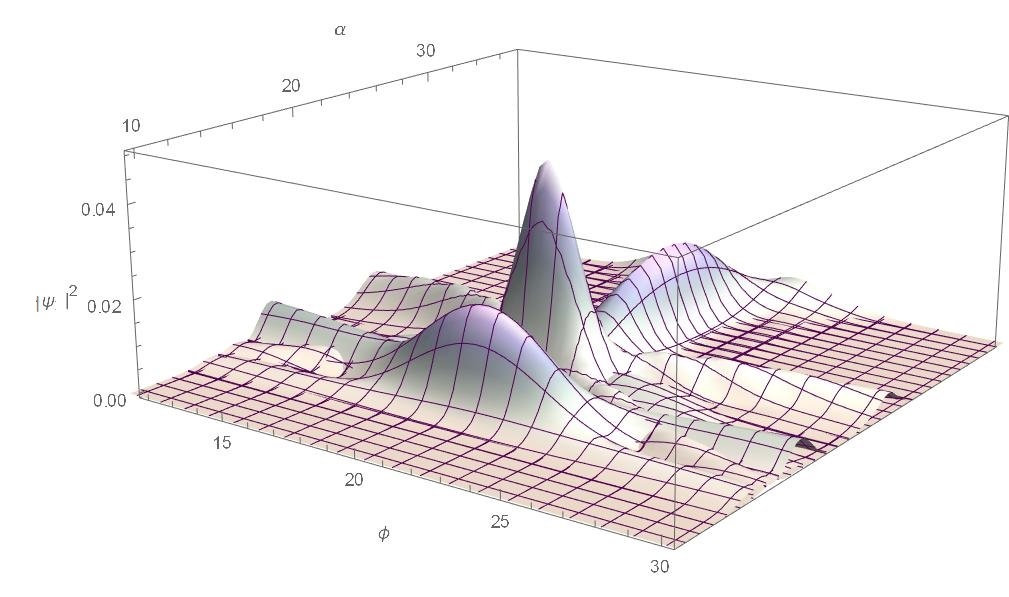} 
\caption{Evolution of the probability density $|\Psi|^2$ for different values of $\xi=10, 1, 0.01$. We take $\bar{k}_{\alpha}=\bar{k}_{\phi}=0.3$ and $\sigma_{\alpha}=\sigma_{\phi}=2$.}
\label{sovrapp}
\end{figure}

Then, we analyse the behaviour of the following wave packet
\begin{widetext}
\begin{equation}
     \Psi=\int^{+\infty}_{-\infty}dk_{\alpha} dk_{\phi}A(k_{\alpha},k_{\phi})e^{i(k_{\alpha}\alpha+ k_{\phi}\phi)}\xi^{i k_{\alpha}} e^{2 i\sqrt{k_{\alpha}^2 - \frac{3c^4}{8\pi G} \xi k_{\phi}^2}} \ e^{-2i k_{\alpha}   arctanh{ \Big(  \frac{\sqrt{k_{\alpha}^2 - \frac{3c^4}{8\pi G}\xi k_{\phi}^2}}{k_{\alpha}}} \Big) }
    \label{psiquadro}
\end{equation}
\end{widetext}
with $A(k_{\alpha}, k_{\phi})$ being the product of two Gaussian functions in $k_{\alpha}$ and $k_{\phi}$ respectively. The probability density $\Psi \Psi^*$ is plotted in Fig.(\ref{sovrapp}).
 We observe that in a model in which we use as time not the matter scalar field but the internal $f(R)$ scalar mode non-minimally coupled to gravity, the universe tends to classicize as the function of this variable time up to a critical value. When the wave function is peaked, one can consider the mean values of $k_{\alpha}$ and $k_{\phi}$ as corresponding to their classical ones, in fact the critical value is $\xi=0.3$ for fixed $\bar{k}_{\alpha}=\bar{k}_{\phi}=0.3$. 
After that, a change of regime in the solution is present, \emph{i.e.} a regime transition of the square root from imaginary to real values, and a subsequent spreading of the wave packet in the $\alpha$ variable arises (when estimated by the mean values), Fig(\ref{sovrapp}).

\begin{figure}[h!] 
\centering
\includegraphics[width=8.5cm]{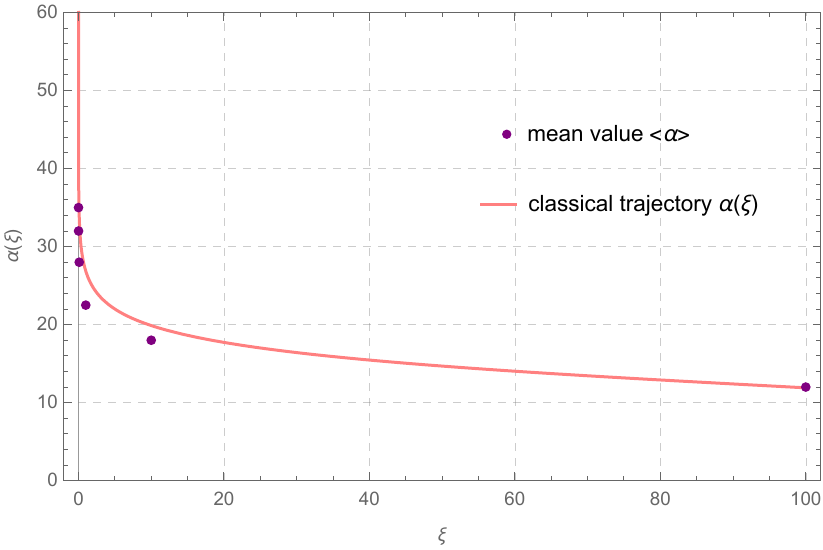} 
\caption{Comparison  between  the  classical  trajectory  (solid line) and the $\alpha(\xi)$ mean values (points).}
\label{traiettoriaesatto}
\end{figure}

To demonstrate the consistency of the classical limit
of the examined quantum model, we compare the classical trajectory provided by the
Hamilton equations (\ref{alphacampo}) and the mean values of the coordinate $\alpha$ as $\xi$ varies. Such a choice of the mean values is to be considered as a more quantitative analysis than the previous one in Sec.(\ref{IV}) and Sec.(\ref{subsecA}). In fact, from the mean values
\begin{equation}
    \langle \alpha \rangle= \int d\alpha d\phi \ \Psi^*\alpha \Psi ,
\end{equation}
we can study the evolution of the basic variable $\alpha$ fluctuation, \textit{i.e}
\begin{equation}
    \Sigma_{\alpha}=\sqrt{\langle \alpha^2 \rangle- \langle \alpha \rangle^2}.
\end{equation}

\begin{figure}[h!] 
\centering
\includegraphics[width=8.5cm]{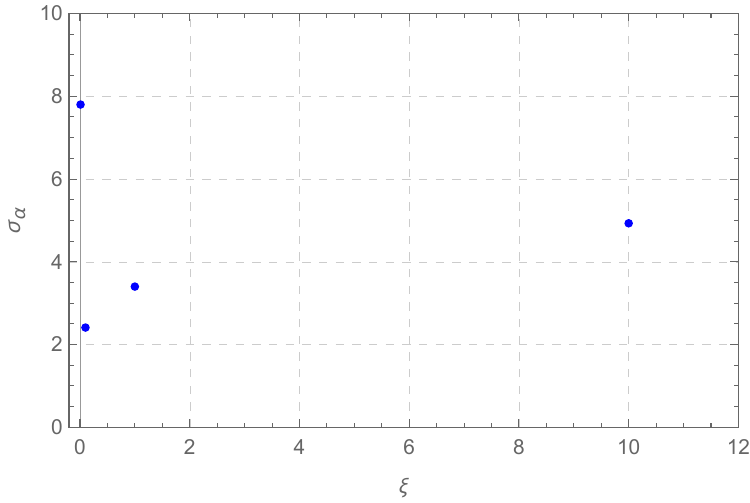} 
\caption{Evolution of the standard deviation $\Sigma_{\alpha}$ at different values of $\xi$. $\Sigma_{\alpha}=\ 7.8,\ 2.41,\ 3.4,\ 4.93$, $\xi=\ 0.01,\ 0.1,\ 1,\ 10$ respectively.}
\label{sigma}
\end{figure}

Where $\langle .. \rangle$ denotes the mean values of the enclosed quantities. Here as well, from Fig.(\ref{traiettoriaesatto}) it can be realized that the quantum expectation value dynamics follows the classical one of the FLRW universe in the \emph{Jordan frame}. Moreover, from Fig.(\ref{sigma}) it is evident the decreasing behaviour of the standard deviation $\Sigma_{\alpha}$ up to the critical value of the time $\xi$ after which it starts further to increase.
It is worth noting that it is reliable the possibility to characterize the quantum probability distribution in terms of $\langle \alpha \rangle$ and $\Sigma_{\alpha}$ because we are dealing with Gaussian wave packets.
It is also the Gaussian nature of our localization that ensures that the quantities $\sigma_{\alpha}$ and $\Sigma_{\alpha}$ naturally obey the uncertainty principle. Our analysis acquires a rather general character, as soon as we consider that a localized wave packet (symmetric with respect to the mean value) can be always approximated by a Gaussian one to a given order.

\subsection{The role of inflation}
The result of the previous section leads to the very surprising feature in the behaviour of the universe volume that an increasing localization of the quantum wave packet takes place as it expands. We clarified how this classicalization tendency of the quantum dynamics is sensitive to the sign of the square root present in the reduced Hamiltonian (\ref{hximatter}).
The idea of a small initial value of $p_{\phi}$ is supported by the scenario of an inflation field that is laying nearby the minimum of its potential, before the phase transition of the inflation paradigm takes place.
However, the possibility for a non-negligible kinetic term of the field $\phi$, near enough the Planckian era can not be ruled out at all from the problem. In fact, the kinetic energy of a massless free scalar field increases toward the singularity as $e^{-3\alpha}$ and this fact is at the ground of the implementation of this term to construct a relational clock for the quantum universe dynamics. Thus, the question concerning a subsequent spreading of the wave packet, after its localization in the variable $\alpha$ must be discussed.

Independently of the mean value taken by $p_{\phi}$ during the Planckian era, we suggest the following scenario to connect the primordial quantum phase with a classical inflationary stage.
During the classicalization process, we can infer that the phase transition of the inflaton paradigm can start, here modelled via the introduction of a cosmological constant term in the (emerging) classical dynamics. This hypothesis is clearly more reliable as smaller is the mean value of $p_{\phi}$ in the quantum phase, since the transition has more available time to take place.

The Hamilton equations describing the universe evolution in the presence of a cosmological constant take the form
\begin{equation}
   \begin{cases}
   \frac{\partial \alpha}{\partial \xi}=-\frac{3\Lambda e^{ \alpha(\xi)}}{\sqrt{p_{\alpha}^2-2\xi(\frac{3c^4}{8 \pi G}p_{\phi}^2+\Lambda e^{3\alpha(\xi)}})}\\
   \frac{\partial p_{\alpha}}{\partial \xi}=\frac{1}{\xi}\biggl(-1-\frac{2p_{\alpha}}{\sqrt{p_{\alpha}^2-2\xi(\frac{3c^4}{8 \pi G}p_{\phi}^2+\Lambda e^{3\alpha(\xi)}})}\biggl)
   \end{cases} 
\end{equation}

We see that for positive value of $p_{\alpha}$ the derivative $\partial \alpha/\partial \xi$ is negative, \emph{i.e.} as without the cosmological constant.
Thus, the universe expands as the clock coordinate $\xi$ decreases (\emph{i.e.} the arrow of time is reversed with respect to the synchronous gauge) and it is worth observing that the critical value of $\xi$ at which without the cosmological constant a singularity behaviour emerged, is instead now regular.

A singular behaviour of the dynamics is now recovered for a smaller value of $\xi$, namely when the following condition takes place
\begin{equation}
    \xi=\frac{p_{\alpha}^2}{2(\frac{3c^4}{8\pi G}p_{\phi}^2+\Lambda e^{3\alpha})}.
    \label{lmx}
\end{equation}

It is very easy to realize that, in correspondence to this singular instant, the quantity $\partial \alpha/\partial \xi$ approaches $-\infty$ and an asymptote for the function $\alpha(\xi)$ arises. Such an asymptotic behaviour as $\xi$ approaches the critical value (\ref{lmx}), contains all the universe inflationary (de Sitter-like) evolution, \emph{i.e.}, in the limit of our representation, all the slow-rolling phase.

These considerations suggest that the presence of a cosmological constant can drive the localized universe volume to a rapid expansion phase, without the possibility of a new spreading of the wave function.

\section{Concluding remarks}\label{VI}

We analyzed the canonical quantum dynamics of the isotropic universe, as viewed in the metric $f(R)$ in the \emph{Jordan frame} and in the presence of a free massless scalar field (thought as the inflaton field in the pre-inflation age).

The peculiarity of our study consists of adopting as time variable the non-minimally coupled scalar field emerging from the scalar-tensor formulation of the $f(R)$ gravity.
We have chosen such a scalar field as a clock for our quantum cosmology before quantizing, i.e. we fix the time gauge which selected this time coordinate and then we solved the Hamiltonian constraint with respect to its conjugate momentum.
As result of this Hamiltonian reduction procedure, we arrive to a Schr\"odinger dynamics for the universe wavefunction which depends on time and on the two space-like variables corresponding to $\alpha$ and $\phi$.

We firstly analysed the vacuum case, by showing the non-spreading character of the localized wave packets up to the singularity. Then, we included the massless scalar field into the dynamics (initially by a perturbation approach and eventually in its general formulation). We demonstrate how the presence of such a scalar field affects the wave packet evolution by inducing a progressive peaking of the wave function, as the universe volume increases (\emph{i.e.} still for increasing values of $\alpha$).

Actually, the progressive localization of the wave function takes place (in the case of generic initial condition for the inflaton) only up to a given instant of time, depending on the Cauchy problem for the Schr\"odinger dynamics.

This peculiar feature suggested a spontaneous mechanism of classicalization, absent in the canonical quantization of the standard Einsteinian picture.
The scenario we infer here is that before a new spreading of the wave packet takes place in the universe future, the inflation phase transition starts, inducing a new dynamical regime. In other words, the universe tends to a classical stage as a natural consequence of its quantum evolution and then it starts to feel the presence of a huge cosmological constant term which maintains this classical behavior and causes a very rapid expansion.

This result calls attention, in view of its generalization to more complicated situations, like the evolution of Bianchi universes \cite{bkl70,Primordial}, \emph{e.g.} the types I, V and IX models, which
generalize to the anisotropic sector the evolution of the three isotropic available geometries \cite{LL}.
It is a well-know result, see \cite{kirmont97} that it is not possible, in the canonical picture, that the universe approaches a classical regime before the oscillations of the Bianchi IX model ends (\emph{i.e.} the so-called Mixmaster model) \cite{misner68}.
For isotropization mechanisms in the picture of a WKB approach for the anisotropy degrees of freedom see \cite{Belvedere,noi,Roberta e Valerio}.

The Bianchi IX model is of particular interest because it admits the isotropic limit and provides a valuable \emph{minisuperspace} model for the generic inhomogeneous universe \cite{Benini2008,l'ultimo con Roberta}.
Thus, the present study suggests that the proposed dynamical paradigm must be extended to more general cosmological models, in order to understand if,
adopting the non-minimally coupled scalar field of the $f(R)$ gravity as physical clock, allows a tendency of
the universe volume to a classical limit, even in the presence of anisotropy degrees of freedom. Such a point of view could justify the implementation of a WKB model for the anisotropy variables, as a natural result of the quantization of the considered models in the context of an extended metric $f(R)$ model, as described in the \emph{Jordan frame}. The idea that the result here obtained survives also in the presence of anisotropies of the universe is made very reliable because of the equivalence that a massless scalar field and the kinetic term associated to the anisotropy contributions single out in the morphology of the gravitational Hamiltonian. Nonetheless, the peculiarity of the potential term associated to this anisotropies, \emph{i.e.} the three-scalar curvature of the universe, can have a significant role in this respect and it calls attention for further investigations.

\end{document}